\documentclass{article}
\usepackage{bre05}
\usepackage{graphicx}
\frompage{000} \topage{000}                                              %|
\def\rightmark{Angular Moment Analysis of Correlations}
\def\leftmark{P.~Danielewicz and S.~Pratt}

\title{Angular Moment Analysis of\\ Low
Relative Velocity Correlations}
\authors{
{P.~Danielewicz$^{1,2,a}$ and S.~Pratt$^{2,b}$ %
}\\[2.812mm]
{\normalsize
\hspace*{-8pt}$^1$ National Superconducting Cyclotron Laboratory and\\[0.2ex]
\hspace*{-8pt}$^2$ Department of Physics and Astronomy, Michigan State University,\\
East Lansing, MI 48824-1321, USA
}}

\abstract{For analyzing anisotropic low relative-velocity correlation-functions
and the associated emission sources, we propose an expansion in terms of
cartesian spherical harmonics.  The expansion coefficients represent angular
moments of the investigated functions.  The respective coefficients for the correlation
and source are directly related to each other via one-dimensional
integral transforms.  The shape features of the source may be partly read off
from the respective features of the correlation function and can be, otherwise, imaged.}


\begin{document}
%\sloppy

\maketitle
\setcounter{page}{1}

\section{Introduction}\label{intro}

Use of interferometry for the determination of reaction geometry
came to nuclear physics from astronomy.  In both fields, the intensity technique has
been first used for determining gross sizes of the radiating regions.  While astronomy has since moved
on to a systematic model-independent determination of the fine details of observed objects \cite{mon03},
within the phase version of the method,
the nuclear physics has not advanced beyond fitting the experimental data
with models that parameterize coarse
shape-features of particle-emitting regions \cite{adl01,ghe03}.  Imprinted in the shape features
can be the reaction history and, specifically, differences in emission times for
different particles and stalling of a system when passing through the quark-gluon
phase transition.  This talk is dedicated to the systematic determination of the
shape features of emitting regions in reactions with multi-particle final states.

\section{Low-Velocity Correlations in Reactions}

Ability to learn on geometry of emitting regions in reactions relies on the possibility of
factorizing the amplitude for the reaction into a wavefunction $\Phi_{\bf q}^{(+)}$ for the pair of detected
particles and an amplitude remnant.  The wavefunction can be computed and, at at low relative velocity,
its square may exhibit pronounced spatial features such as associated with identity interference,
resonances or
Coulomb repulsion.  Those features can be regulated by changing the relative particle
momentum~$q$.  The amplitude
remnant, squared within a cross section and summed over the
unobserved particles and integrated over their momenta, yields $S'$ which
represents general features of reaction geometry, without a significant variation with~$q$.
When looking at the inclusive two-particle cross-section then,
the structures in $|\Phi^{(+)}|^2$ can be used to explore the geometry in $S'$:
\begin{equation}
\frac{d \sigma}{d{\bf p}_1 \, d{\bf p}_2} =  \int  {\rm d} {\bf r} \,  S'_{\bf P} ({\bf r}) \,
 |\Phi^{(-)}_{{\bf q}} ( {\bf r})|^2 \, .
\end{equation}

The naive expectation, met at large $q$ for a multi-particle final state, is that the emission
of two particles is uncorrelated.  Thus, one normalizes the two-particle cross section with
a product of single-particle cross sections, and looks for the deviations of the thus constructed
correlation function $C$ from 1:
\begin{equation}
{\mathcal R}({\bf q}) =
C({\bf q}) - 1 = \frac{\frac{1}{\sigma} \, \frac{d \sigma }{ d{\bf p}_1 \, d{\bf p}_2 } }
{\frac{1}{\sigma} \, \frac{d \sigma}{d{\bf p}_1 }   \, \frac{1}{\sigma}  \, \frac{d \sigma}{d{\bf p}_2 }}
- 1 = \int  {\rm d} {\bf r} \,
 \left( |\Phi^{(-)}_{{\bf q}} ( {\bf r})|^2 - 1 \right) \,
 S_{\bf P} ({\bf r}) \, .
 \label{eq:R}
\end{equation}
The last equality follows from the fact that, at large $q$, the relative wavefunction
squared is equal, on the average, to~1; in combining this with $C \simeq 1$ at large $q$,
the source $S$, following
the cross-section normalization, turns out to be normalized to~1, $\int {\rm d} {\bf r} \, S({\bf r}) = 1$.
Equation (\ref{eq:R}) links the deviations of correlation function $C$ from~1,
to the interplay of deviations of the relative wavefunction from~1 with the geometry in~$S$.
With $S$ normalized to 1, it may be interpreted as the probability of emitting the two
particles at the relative separation~${\bf r}$ within the particle center of mass.
The emission is integrated over time, as far as $S$ is concerned.

\section{Analysis of Correlations}

Knowing ${\mathcal R}({\bf q})$ and $|\Phi|$, on can try to learn about
$S$; mathematically, this represents a difficult problem involving an
inversion of the integral kernel $K({\bf q}, {\bf r}) = |\Phi^{(-)}_{{\bf
q}} ( {\bf r})|^2 - 1$, in three dimensions.  The situation is simplified by the fact
that $K$ depends only on $q$, $r$ and the angle $\theta_{\bf q \, r}$, and, correspondingly,
can be expanded in Legendre polynomials, $K({\bf q},{\bf r}) =
\sum_\ell (2 \ell +1 )\, K_\ell (q,r) \, P^\ell
(\cos{\theta})$.  If the correlation and source functions are expanded in spherical harmonics
$Y^{\ell m}$, ${\mathcal R} ({\bf q}) = \sqrt{4 \pi} \sum_{\ell
m} {\mathcal R}^{\ell m} (q) \, {\rm Y}^{\ell m} (\hat{\bf q})$, one finds that
the three-dimensional relation (\ref{eq:R}) is equivalent to a set of one-dimensional
relations for the harmonic coefficients \cite{bro97}
\begin{equation}
{\mathcal R}^{\ell m}(q)
= 4 \pi \int {\rm d}r \, r^2 \, K_\ell (q, r) \,
S^{\ell m} (r) \, .
 \label{eq:Rlm}
\end{equation}
For weak anisotropies, only low-$\ell$ coefficients of ${\mathcal R}$ or $S$ are expected to be
significant.  The $\ell=0$ version of (\ref{eq:Rlm}) connects the angle-averaged functions.

The above suggests a systematic analysis of the correlation functions and source
functions in
terms of the harmonic coefficients.  A problem, however, is that it is cumbersome to analyze
real functions in terms of complex coefficients lacking a clear interpretation for $\ell \ge 1$.
This suggests looking for another basis for the directional decomposition of
correlation functions and sources, that would be real and have a a clear geometric meaning.  Such a basis may
be constructed starting from a unit direction vector
$\hat{n}_\alpha = (\sin{ \theta} \cos{\phi}, \sin{ \theta} \sin{\phi}, \cos{ \theta})$.

The tensor
product of $\ell$ vectors $\hat{n}_\alpha$ yields a symmetric rank-$\ell$ cartesian tensor that is a
combination of spherical tensors of rank $\ell' \le \ell$ and the same evenness as~$\ell$:
\begin{equation}
(\hat{n}^{\ell})_{\alpha_1 \ldots \alpha_{\ell}} \equiv \hat{n}_{\alpha_1} \, \hat{n}_{\alpha_1} \ldots \hat{n}_{\alpha_\ell} =
\sum_{\ell' \le \ell, m} c_{\ell' m} \, {\rm Y}^{\ell' m} \, .
\end{equation}
A projection operator $\mathcal{D}$ may be constructed in the space of symmetric
rank-$\ell$ tensors, out of a combination of Kronecker $\delta$-symbols, that makes a symmetric
tensor traceless,
\begin{equation}
\sum_{\alpha} ({\mathcal D} \hat{n}^\ell)_{\alpha \, \alpha \, \alpha_3 \ldots \alpha_\ell} = 0
\, .
\end{equation}
The tracelessness of $({\mathcal D} \hat{n}^\ell)$ ensures that the products
$r^\ell \, ({\mathcal D} \hat{n}^\ell)_{\alpha_1 \ldots \alpha_\ell}$
are solutions of the Laplace equation and, thus, the
$({\mathcal D} \hat{n}^\ell)_{\alpha_1 \ldots \alpha_\ell}$ are combinations of spherical
harmonics of rank $\ell$ only.  The components are real and may be used to
replace~$Y^{\ell m}$.  The lowest-rank tensors are:
\begin{eqnarray}
{\mathcal D}\hat{n}^{0}  =  1 \, , \hspace*{1.5em}
({\mathcal D}\hat{n}^{1})_\alpha  =  \hat{n}_\alpha \, , \hspace*{1.5em}
({\mathcal D}\hat{n}^{2})_{\alpha_1 \, \alpha_2}  =   \hat{n}_{\alpha_1} \,  \hat{n}_{\alpha_2}
- \frac{1}{3} \delta_{\alpha_1 \, \alpha_2} \, , \nonumber \\
({\mathcal D} \hat{n}^{3})_{\alpha_1 \,
\alpha_2 \, \alpha_3}  =  \hat{n}_{\alpha_1} \,  \hat{n}_{\alpha_2} \,  \hat{n}_{\alpha_3}
- \frac{1}{5} ( \delta_{\alpha_1 \, \alpha_2} \, \hat{n}_{\alpha_3}
+ \delta_{\alpha_1 \, \alpha_3} \, \hat{n}_{\alpha_2} + \delta_{\alpha_2 \, \alpha_3} \, \hat{n}_{\alpha_1}) \, .
\end{eqnarray}

The completeness relation in terms of the cartesian components is \cite{dan05}
\begin{eqnarray}
\delta(\Omega' - \Omega) & =  & \frac{1}{4 \pi} \sum_{\ell} \frac{(2\ell+1)!!}{\ell !}
\sum_{\alpha_1 \ldots \alpha_\ell}
({\mathcal D}\hat{n'}^{\ell})  _{\alpha_1 \ldots \alpha_{\ell}} \,
({\mathcal D}\hat{n}^{\ell}) _{\alpha_1 \ldots \alpha_{\ell}} \nonumber \\
& = & \frac{1}{4 \pi} \sum_{\ell} \frac{(2\ell+1)!!}{\ell !}
\sum_{\alpha_1 \ldots \alpha_\ell}
({\mathcal D}\hat{n'}^{\ell})  _{\alpha_1 \ldots \alpha_{\ell}} \,
\hat{n}_{\alpha_1} \ldots \hat{ n}_{\alpha_{\ell}} \, ,
\end{eqnarray}
where the second equality follows from ${\mathcal D}={\mathcal D}^\top={\mathcal D}^2$.
The completeness relation can be used for expanding ${\mathcal R}$ or $S$ in terms of
cartesian tensor components
\begin{equation}
{\mathcal R}({\bf q})= \int {\rm d}\Omega' \, \delta(\Omega' - \Omega) \, {\mathcal R}({\bf q}') = \sum_\ell \sum_{\alpha_1 \ldots \alpha_\ell}
{\mathcal R}_{\alpha_1 \ldots \alpha_\ell}^{(\ell)}(q) \, \hat{ q}_{\alpha_1} \ldots \hat{ q}_{\alpha_\ell}
\, ,
\end{equation}
where the coefficients are angular moments,
${\mathcal R}_{\alpha_1 \ldots \alpha_\ell}^{(\ell)} (q) = \frac{(2\ell+1)!!}{\ell !} \int \frac {{\rm d} \Omega_{\bf q}}{4 \pi} \,
{\mathcal R}({\bf q}) \, ({\mathcal D}\hat{q}^{\ell}) _{\alpha_1 \ldots \alpha_{\ell}}$.

With the ${\mathcal R}$ and $S$ cartesian coefficients being identical combinations
of the respective $Y^{\ell m}$-coefficients, the corresponding ${\mathcal R}$ and $S$
cartesian coefficients are directly related to each other,
\begin{equation}
{\mathcal R}^{(\ell)}_{\alpha_1\cdots\alpha_\ell}(q)
= 4 \pi \int  {\rm d}r \, r^2 \,
K_\ell(q,r) \,
{\mathcal S}^{(\ell)}_{\alpha_1\cdots\alpha_\ell}(r) \, .
\end{equation}
For weak anisotropies, only low $\ell$-coefficients for the source or correlation matter,
\begin{equation}
{\mathcal R}({\bf q})  =
{\mathcal R}^{(0)}(q) + \sum_\alpha \, {\mathcal R}^{(1)}_\alpha(q) \,
\hat{q}_\alpha + \sum_{\alpha_1 \, \alpha_2}
{\mathcal R}^{(2)}_{\alpha_1 \alpha_2} (q) \,
\hat{q}_{\alpha_1} \, \hat{q}_{\alpha_2} + \ldots
\, .
\end{equation}
Here, ${\mathcal R}^{(0)}$ is the angle-averaged function.
The dipole and quadrupole distortions, ${\mathcal R}^{(1)}$ and ${\mathcal R}^{(2)}$,
can be represented in terms of amplitudes and distortion
vectors: ${\mathcal R}_{\alpha}^{(1)} = R^{(1)} \, e_{\alpha}^{(1)}$ and
${\mathcal R}^{(2)}_{\alpha \beta} (q) =R_1^{(2)} \,
e_{1 \alpha}^{(2)} \, e_{1 \beta}^{(2)} + R_3^{(2)} \, e_{3 \alpha}^{(2)} \,
e_{3 \beta}^{(2)} - \left( R_1^{(2)} + R_3^{(2)} \right) \, e_{2 \alpha}^{(2)}
\, e_{2 \beta}^{(2)}$, respectively.

\section{Illustration}

\begin{figure}[htb]
\parbox{.45\linewidth}{
\includegraphics[width=.94\linewidth]{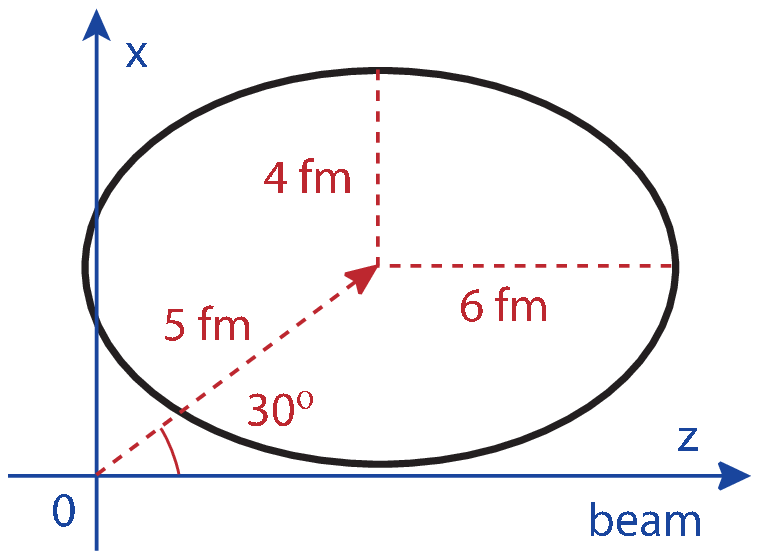}
}
\parbox{.55\linewidth}{\hspace*{-2em}
\includegraphics[width=1.1\linewidth,height=.75\linewidth]{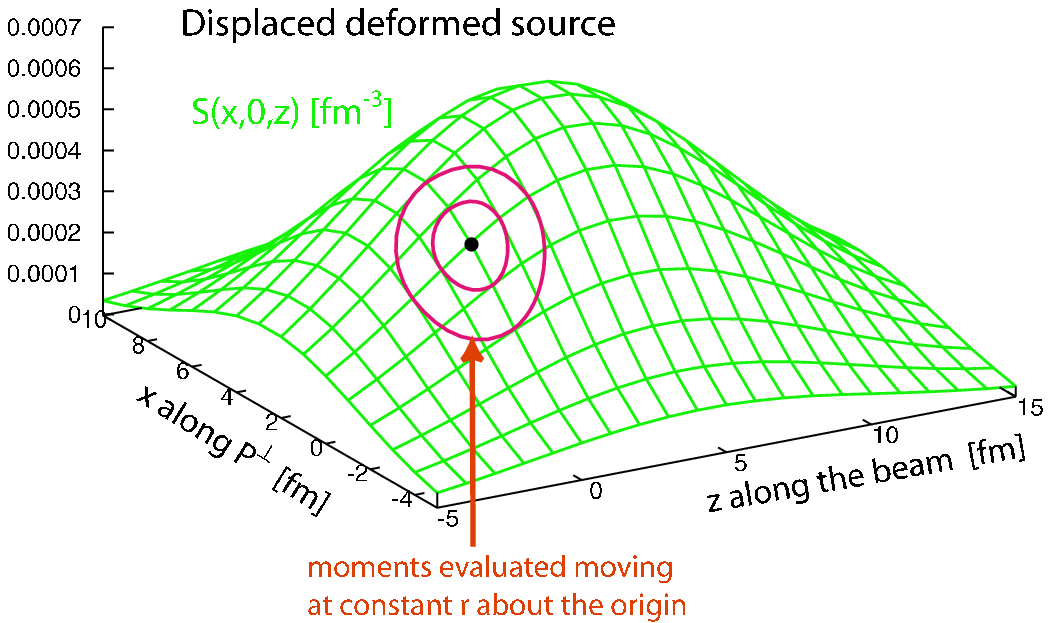}
}
\vspace*{-.4cm}
\caption[]{Sample Gaussian emission source within the plane of relative
directions, with the $z$-axis pointing along the beam and the $x$-axis
along ${\bf P}^\perp$.
}
\label{fig:source}
\end{figure}

\begin{figure}[htb]
\parbox{.5\linewidth}{
\includegraphics[width=.9\linewidth]{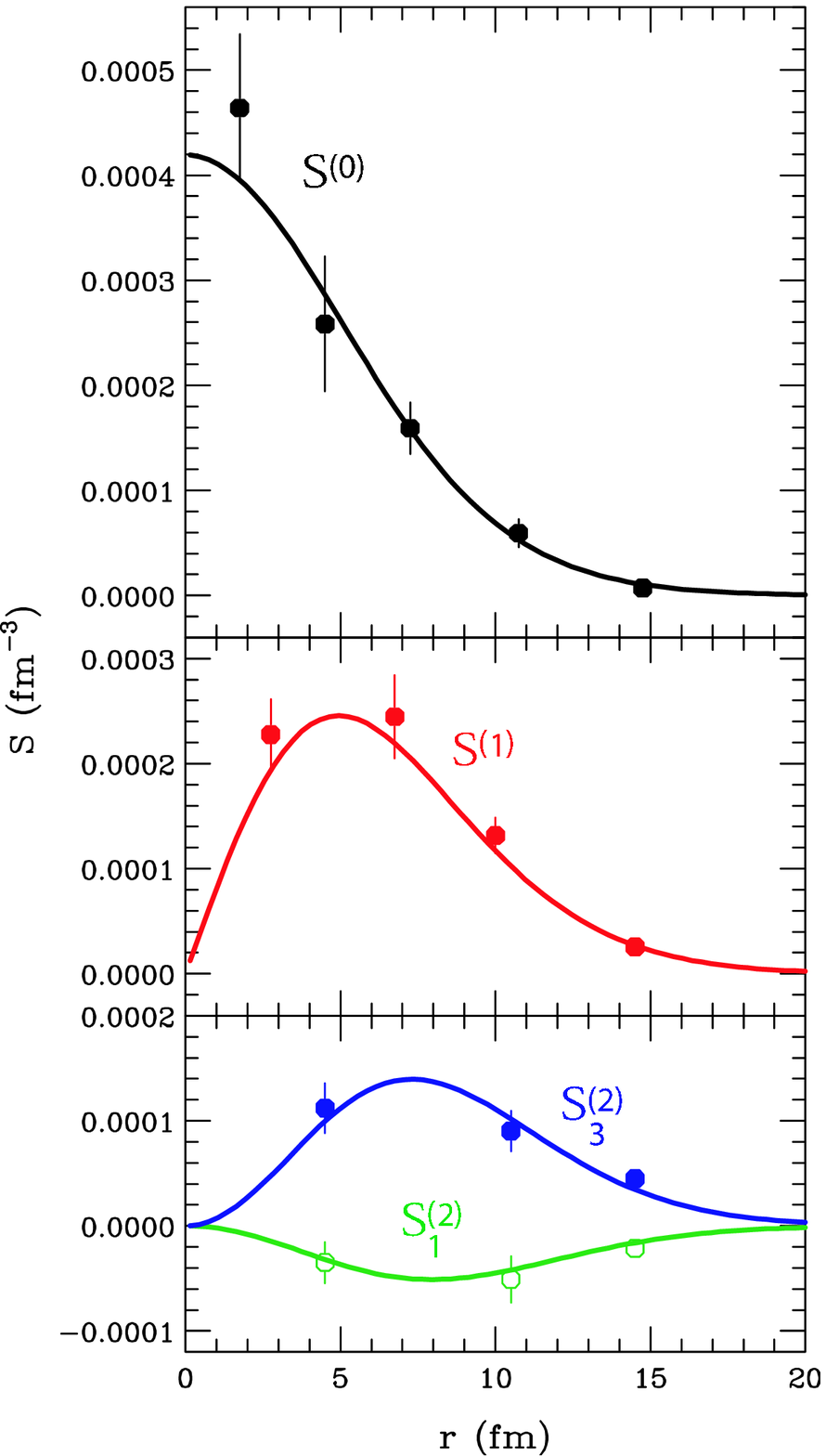}
}
\parbox{.5\linewidth}{
\includegraphics[width=.9\linewidth]{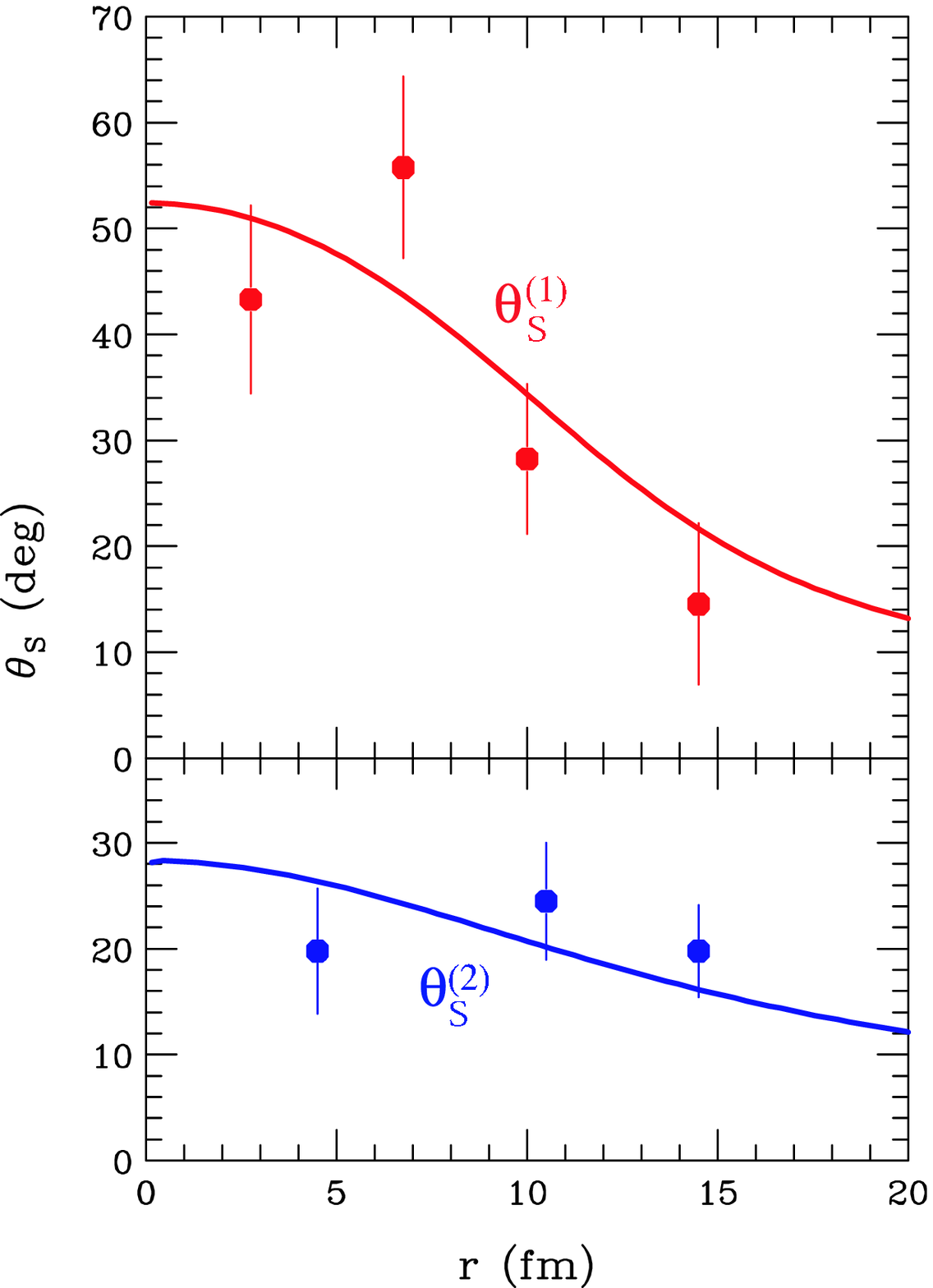}
}
\vspace*{-.4cm}
\caption[]{Low-$\ell$ values (left) and angles (right) for the relative source,
as a function of the relative distance $r$.  Lines and symbols represent, respectively, the exact
values and values obtained through imaging of the anisotropic correlation function.
}
\label{fig:sourcemo}
\end{figure}

We illustrate the analysis in terms of cartesian harmonics taking an anisotropic
Gaussian source, elongated along the beam axis, and displaced along the total
pair momentum ${\bf P}$, at an angle of 30$^\circ$ relative to the beam axis, cf.\
Fig.\ \ref{fig:source}.
Figure \ref{fig:sourcemo} shows low-$\ell$ source values and directions,
from the sample source decomposition in terms of the cartesian harmonics.  The angle-average
source maximizes at $r=0$, although the original source is displaced from the origin.  The
distortions can only become significant at distances comparable to source size; in fact,
at small distances $S^{(\ell)} \propto r^\ell$.

For the anisotropic source, we next construct a correlation function,
assuming the classical limit of repulsive Coulomb interactions, suitable for emitted
intermediate mass fragments.  In that limit, the kernel $K$ is a function of
of $\theta_{\bf q r}$ and $r/r_c$, where $r_c$ is the distance of closest approach in
a head-on collision, $\frac{q^2}{2m_{ab}}= \frac{Z_a \, Z_b \, e^2}{4\pi
\epsilon_0 \, r_c}$, as
\begin{equation}
|\phi|^2 = \frac{{\rm d}^3 \, q_0}{{\rm d}^3 \, q} =
\frac{\Theta \left( 1 + \cos{\theta_{{\bf q} {\bf r}}}-{2 r_c}/{r} \right) \,
(1 + \cos{\theta_{{\bf q} {\bf r}}}-{r_c}/{r})}
{\sqrt{\left(1 + \cos{\theta_{{\bf q} {\bf r}}}\right)^2-
\left(1 + \cos{\theta_{{\bf q} {\bf r}}}\right)
{2 r_c}/{r} } } \, .
\end{equation}
The $\ell=0$ kernel is $K_0 = \Theta(r -r_c) \, \sqrt{1-r_c/r} - 1$, while the higher-rank
kernels may be obtained numerically.

\begin{figure}[htb]
\begin{center}
\includegraphics[width=.4\linewidth]{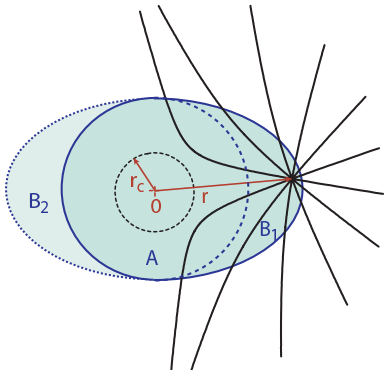}
\end{center}
\vspace*{-.8cm}
\caption[]{
A schematic relative source with repulsive Coulomb trajectories superimposed coming
out isotropically from position ${\bf r}$ within the source.  The region of radius $r_c$
is inaccessible to the trajectories.
}
\label{fig:traj}
\end{figure}

In the classical limit, the correlation function reflects the distribution of
relative trajectories emerging from an anisotropic source, cf.\ Fig.\ \ref{fig:traj}.
In the low-energy limit of $r-r_c \ll r_c$, the trajectories turn back away
from $r=0$, lumping around the direction of ${\bf r}$ they emerge from.  For $r_c$ comparable
to the source size, the
low-energy limit holds and only the source margins contribute to the correlation.
In the high-energy limit of $r_c \ll r$, the trajectories emerge isotropically,
except for the shadow left by the $r < r_c$ region in the $-{\bf r}$ direction.
For $r_c$ small compared to the source size, the high-energy limit generally holds and
the correlation represents the integral features
of the source, as nearly all source points contribute.

\begin{figure}[htb]
\begin{center}
\includegraphics[width=.56\linewidth]{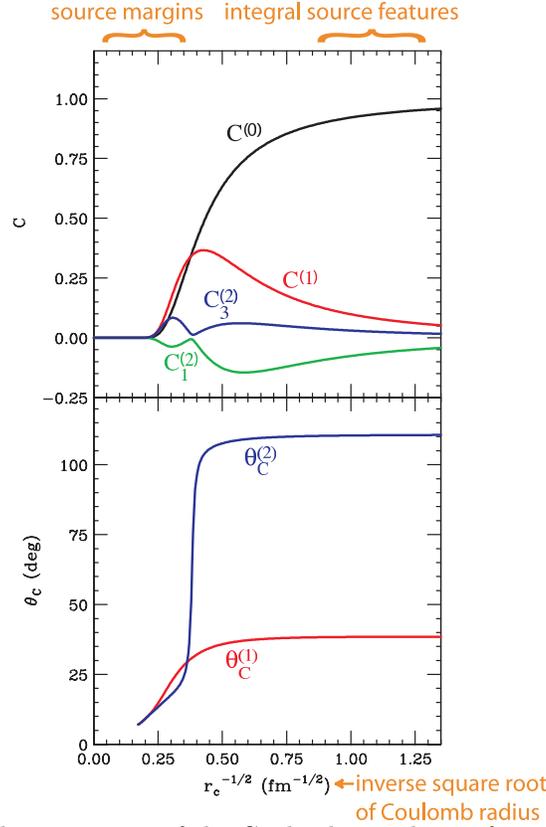}
\end{center}
\vspace*{-.8cm}
\caption[]{
Low-$\ell$ characteristics of the Coulomb correlation function for our sample source:
amplitudes (top panel) and angles (bottom panel) as a function of the inverse square
root of the head-on return radius $r_c$.
}
\label{fig:cor}
\end{figure}

Characteristics of the classical Coulomb correlation function for the source
in Figs.\ \ref{fig:source} and \ref{fig:sourcemo} are next shown in Fig.\ \ref{fig:cor},
plotted there vs.\ $r_c^{-1/2} \propto q$.  The dipole angle for $C$ follows the variation
of the dipole angle for $S$.  On the other, the quadrupole angle exhibits a jump by 90$^\circ$
as a function of $r_c^{-1/2}$.  This jump is associated with the change in sign for $K_2$ and the resulting
prolate-to-oblate shape transition for $C$.  For more schematic sources, one or more
low-$\ell$ correlation amplitudes may vanish and/or angles may exhibit less
variation.

We further test out the imaging \cite{bro97,bro98} of the features of our anisotropic source from
the correlation function.  We assume that the $\ell \le 2$ cartesian coefficients
have been measured for our Coulomb correlation function, at 80 values of $r_c^{-1/2}$.
We generate the coefficient values randomly assuming a constant r.m.s.\ error of~0.015~\cite{dan05}.
From the simulated measurements, we restore the coefficients for the source, within
the region
of $r < 17$~fm, at the number of points, in $r$, that drops with~$\ell$: 5~for $\ell=0$,
4 for $\ell=1$ and 3
for $\ell = 2$.  The results of restoration, represented by symbols in
Fig.\ \ref{fig:sourcemo}, compare
reasonably with the original source features.

Given the cartesian coefficients for the source as a function of $r$, it is straightforward
to compute the cartesian moments of the source.  The moments from imaging are compared
in Table \ref{tab:moments} to those computed directly and
they are found to reproduce the latter reasonably.
\begin{table}[hb]
\vspace*{-12pt}
\caption[]{Cartesian moments and weight of the source at $r < 17$~fm,
obtained in imaging and by direct integration.
}\label{tab:moments}
\vspace*{-14pt}
\begin{center}
\begin{tabular}{r| c |r @{$\pm$} l |c|}
Quantity & Unit & \multicolumn{2}{c|}{Restored} & Original \\
\hline
$4\pi \int {\rm d} r \, r^2 \, S^{(0)}$ & & 0.99 & 0.05 & 1.00\\
$\langle x \rangle$ & fm & 2.47 & 0.11 & 2.45 \\
$\langle z \rangle$ & fm & 4.25 & 0.13 & 3.90 \\
$\langle (x - \langle x \rangle)^2 \rangle^{1/2}$ & fm & 3.80 & 0.24 & 3.90 \\
$\langle y^2 \rangle^{1/2}$ & fm & 3.81 & 0.22 & 3.91 \\
$\langle
(z - \langle z \rangle)^2 \rangle^{1/2}$ & fm & 5.54 & 0.19 & 5.60 \\
$\langle (x - \langle x \rangle )( z - \langle z \rangle ) \rangle$ &
fm$^2$ & 2.23 & 1.49 & -0.41 \\
\hline
\end{tabular}
\end{center}
\end{table}

\section{Conclusions}\label{concl}

Characteristics of the anisotropic correlation functions in reactions, and of corresponding emission sources,
may be quantitatively expressed in terms of the cartesian coefficients that represent
angular moments of the functions.  The coefficients for the correlation
and the source depend on magnitude
of the relative momentum and separation, respectively.  The respective coefficients for the source
and correlation are related to each other through one-dimensional integral transforms.
Certain shape features for the source may be directly read off from the correlation features in terms
of the coefficients.
Otherwise, the source features may be imaged.

\section*{Acknowledgments}

The authors thank David Brown for discussions and for collaboration on a related project.  This work
was supported by the U.S.\ National Science Foundation under Grant PHY-0245009 and by the U.S.\
Department of Energy under Grant No. DE-FG02-03ER41259.

\section*{Notes}
\begin{notes}
\item[a]
E-mail: danielewicz@nscl.msu.edu
\item[b]
E-mail: pratts@pa.msu.edu
\end{notes}

\vfill\eject
\end{document}